\documentclass[12pt]{article}        
\usepackage{setspace}
\usepackage{cite}
\usepackage{amsmath}
\usepackage{parskip}
\usepackage{float}
\usepackage{graphicx}
\usepackage{geometry}
\usepackage{apacite}
\usepackage{longtable}
\usepackage{lscape}
\usepackage{array}
\newcommand{\PreserveBackslash}[1]{\let\temp=\\#1\let\\=\temp}
\newcolumntype{C}[1]{>{\PreserveBackslash\centering}p{#1}}
\newcolumntype{R}[1]{>{\PreserveBackslash\raggedleft}p{#1}}
\newcolumntype{L}[1]{>{\PreserveBackslash\raggedright}p{#1}}
\begin{document}
		
\title{Common Ground In Crisis: Causal Narrative Networks of Public Official Communications During the COVID-19 Pandemic}

\author{\thanks{This work was supported by the Division of Civil, Mechanical and Manufacturing Innovation (CMMI) of the National Science Foundation under Grant \#CMMI-2027475 awarded to Carter T. Butts and Grant \#CMMI-2027399 awarded to Jeannette Sutton. This work was also supported by the Computing and Communication Foundations (CCF) of NSF under Grant \#CCF-2200274.}Sabrina Mai\thanks{Department of Sociology, University of California, Irvine, Irvine, California; \texttt{sabrim2@uci.edu}} \and Scott Leo Renshaw\thanks{Software and Societal Systems Department, Carnegie Mellon University, Pittsburgh, Pennsylvania} \and Jeannette Sutton\thanks{College of Emergency Preparedness Homeland Security and Cybersecurity, University at Albany, State University of New York, Albany, New York} \and Carter T. Butts\thanks{Departments of Sociology, Statistics, Computer Science, and EECS and Institute for Mathematical Behavioral Sciences, University of California Irvine, Irvine, California}}

\maketitle

\abstract{
	
	This study investigates the use of causal narratives in public social media communications by U.S. public agencies over the first fifteen months of the COVID-19 pandemic.  We extract causal narratives in the form of cause/effect pairs from official communications, analyzing the resulting semantic network to understand the structure and dependencies among concepts within agency discourse and the evolution of that discourse over time.  We show that although the semantic network of causally-linked claims is complex and dynamic, there is considerable consistency across agencies in their causal assertions.  We also show that the position of concepts within the structure of causal discourse has a significant impact on message retransmission net of controls, an important engagement outcome.
	
	}

	\section{Introduction}
	
The COVID-19 pandemic has been characterized by an overabundance of information, misinformation, and mixed messages that can be costly to the public in terms of time, funds and, more importantly, preventable deaths \cite{pian_causes_2021}. During such turbulent times, the public in countries like the United States looks toward government officials and public agencies to understand the current situation, its causes, and its ramifications \cite{danzig_effects_1958}, particularly for hazards with which populations have less experience. Public entities in turn provide information in response to public demand, and to advance policy objectives (e.g., responding to the hazard, mitigating losses, and minimizing disruption to economic and other activities). Among the strategies employed by public entities in such communications is the use of \emph{structured narratives} -- indicating the ``who, what, where, and why'' relating to the threat -- to provide easily understandable information. Within the family of structured narratives, perhaps the most important is the \emph{causal narrative}, i.e. narratives that describe events in terms of the conjunction of a putatively active event or entity (a ``cause'') and an event, condition, or circumstance claimed to result or follow from the action of the entity (an ``effect'').  Causal narratives can be viewed as providing a ``story form'' for describing events in the world, a format that has been found to be highly mnemonically available across cultures and age ranges \cite{mandler.et.al:cd:1980} and likely to be attended to.  Causal narratives also supply a highly condensed, structured method of imparting information to audiences -- in the context of the COVID-19 pandemic, potentially relaying essential, clarifying, and potentially life-saving information to a lay audience.  
	
In practice, however, the narratives encountered by the public rarely reflect the product of a single speaker: individuals may encounter many different claims from public entities, whose sources are likely to be imperfectly remembered and lumped together to form a general sense of ``what the government is saying.'' (Among the contributors to this is the poor quality of source memory \cite{fragale.heath:jpsp:2004}, the size and complexity of government bureaucracies (and their public-facing elements), and the fact that many claims are encountered second-hand via transmission from others who first encountered them \cite{richardson.et.al:cjs:1979}.)  This is exacerbated by the rise in modern public communication of \emph{terse messaging}: the use of extremely short, often singular messages that are broadcast to the public \cite{sutton.et.al:ics:2014}.  Terse messaging has become particularly prominent via its role in the social media platforms of the 2010s (and into the present), although terse messages are also employed by many warning and alert systems, the chyrons employed by some television channels, and other short-form media.  Because terse messages are by definition short (often only one or perhaps two sentences in length), they cannot contain complex narrative structure, usually being limited to 1-2 causal units (i.e., ``$A$ because of $B$'' assertions) per message.  On the other hand, the brevity of terse messages and their heavy use in informationally high-traffic settings like social media place a premium on salience and mnemonic availability, attributes at which causal narratives excel.  This motivates a strategy in which selected causal assertions are pushed to the public via social media and other channels, hopefully to be encountered and assembled by members of the public into a sufficiently complete understanding to take appropriate action.  This poses challenges, however, among them the prospect that the welter of messages from different agencies will lead individuals to encounter an aggregate narrative (i.e., \emph{discourse}) that is incomplete, incoherent, or inconsistent (either across types of entities or across time).  Moreover, the properties of the discourse may have an impact on the effectiveness of individual messages themselves, with the position of a causal assertion within the discourse either enhancing or suppressing its plausibility or salience.  In countries like the United States, where communication by public agencies is decentralized and generally uncoordinated, there are few guarantees that these challenges will be well-met.

In this paper, we study this phenomenon in the context of the COVID-19 pandemic, examining causal narratives employed in terse social media messages sent by American public agencies during its first 15 months.  Our data is drawn from a census of communications on the then widely-used microblogging service Twitter by 721 public health, emergency management agencies, and elected official accounts (Governors and Mayors) during the period, as well as associated data on message retransmission.  After identifying causal assertions within these materials, we construct semantic networks which map the dyadic relations between putative causes and effects employed by communicators, subsequently analyzing them to examine the underlying structure and interdependencies of the resulting discourse (as well as differences across actors and over time). Finally, using these extracted narrative and network features, we employ a negative binomial regression model to understand how the use of these various cause and effect narrative strategies are associated with the propensity for messages containing causal assertions to be retransmitted -- a form of engagement that is particularly important in the hazard context.  As we show, the discourse among this group of agencies is extremely complex, but maintains a high level of consistency overall, indicating that even diverse agencies with different missions and operating in different regions were able to develop and communicate a common, consensus view of the pandemic despite public controversies during the period.  We also find the structure of the discourse to be related to message retransmission, suggesting that the ``collective story'' told by the welter of public messages does have an impact on how messages are perceived and engaged with.
	
The remainder of the paper proceeds as follows.  Section~\ref{sec_background} briefly reviews relevant background, while Section~\ref{sec_methods} describes data and methods.  Results are presented in Section~\ref{sec_results}, followed by a discussion of further issues in Section~\ref{sec_discussion}.  Section~\ref{sec_conclusion} concludes the paper.
	
\section{Background} \label{sec_background}

Before proceeding to a discussion of our data and methods, we briefly review key results on causal narratives and terse messaging in hazard settings, which motivate our subsequent analyses.

\subsection{Causal Narratives}
The use of narratives as analytical constructs used to piece together sequences of events in a coherent manner has a long history in the social sciences, particularly within historical sociology \cite{cullagh_colligation_1978,griffin_temporality_1992,griffin_narrative_1993}. Narrative as a communicative process has been defined as ``colligation,'' where individuals identify the interrelationships between individual elements (ideas, facts, individuals, and/or organizations) to serve as sense-making explanations/argumentation \cite{mink_autonomy_1966}. Through a structured narrative description, communicators often articulate causal relations to make sense of events, or argue to others that an entity or event ``caused'' an action as an explanation of its occurrence \cite{abell_syntax_1987}. For the purposes of this study, we focus on causal narratives, which exhibit a particular structural form indicating a causal relationship in which one entity or event variable affects some outcome variable.\footnote{Note that, for purpose of this study, our focus is on \emph{assertions,} made in ``folk'' causal language.  We do not thus distinguish between different notions of causality, nor attempt to assess the validity (or even empirical meaningfulness) of the claims in question; our concern is purely with discourse.} 

 In addition to sense-making, causal narratives are often helpful for other essential cognitive functions such as mnemonic retrieval \cite{schank_knowledge_1995,hastie_rational_2009}, making the event at hand more salient and memorable to the reader. Furthermore, causal narratives are often employed as a tool for framing events, where a particular frame may aid in ``making sense of relative events and \textit{suggesting what is at issue}'' \cite{gamson_changing_1987}. The use of particular causal narratives by officials in their public communications may affect how the public considers the situation at hand \cite{nelson_toward_1997}. Previous studies have also shown that framing effects may impact risk perception and subsequent actions taken in the realms of crisis communications \cite{mcclure_framing_2009} and public health \cite{witte_predicting_1996,kelly_effects_2016}. Although some recent studies have investigated frames utilized and their effects in the context of the COVID-19 pandemic \cite{bolsen_framing_2020,deslatte_shop_2020, dhanani_why_2021, hubner_how_2021}, there has been little work investigating the underlying \textit{structure} of COVID-19 pandemic-related discourse.

Many quantitative techniques have been developed for the task of causal narrative analysis, including more cognitively focused frameworks such as Affect Control Theory, in which social events are represented in terms of a story-like structure involving a perceived actor behaving in a specific way toward a perceived object \cite{robinson_affect_2006}. Several approaches have been formulated with the goal of constructing systematic methods of narrative analysis in which causal relations are induced into graph structures. These methods have included algebraic methods of intentional actions and consequences \cite{abell_syntax_1987}, event-structure modeling \cite{heise_modeling_1989} and Bayesian causal networks \cite{pearl_bayesian_2011} (although the latter are often employed as a model for causal inference, as opposed to a framework for understanding mental models or causal narratives per se). Because narratives with particular structures have been found to affect decision-making by impacting subjective beliefs \cite{spiegler_bayesian_2016,eliaz_model_2020}, this is especially salient to our interests of communication during the COVID-19 pandemic -- particularly the aggregate structure of the discourse created by causal narratives.

\subsection{Official Online Communications During Crises}

During disasters and other unfolding hazard events, human populations look to public officials to understand the evolving context and to obtain updates on event impacts and guidance for protective action \cite{danzig_effects_1958}. In the past two decades, officials (e.g. police officers, politicians)  and organizations (e.g., emergency management and public health agencies, Federal agencies like the National Weather Service, and global entities like the World Health Organization) have increasingly leveraged online channels for communication, including social media platforms such as Twitter and Facebook, to bring their messages to the public. The affordances of these digital platforms necessitates a deeper understanding of their use \cite{hughes_social_2015,schmidt_are_2018, luna_social_2018}.

The widespread (if slow) embrace of social media messaging by public agencies within the developed world (and particularly the United States), and the resulting increase in public-facing communication by such agencies can contribute to the common misconception that such organizations are united, or even on the same page, in terms of their communications strategies and goals. In fact, discourse varies from agency to agency (and often within agencies) -- for instance, public health agencies are more oriented toward protection of public health whereas emergency management agencies are more apt to discuss matters relating to crisis logistics, hazard preparedness and mitigation, and responses to disaster events.  In an event of overarching public interest, such as the COVID-19 pandemic, this creates the risk of inconsistent messaging.  Message consistency, both across time and between communicators, has been found to reduce public uncertainty and increase the credibility of the source communicator \cite{wray_communicating_2008}, while inconsistency has been found to reduce the likelihood that individuals will take protective actions \cite{seeger_best_2006}.  Thus, the potential for decentralized messaging to lead to a ``Tower of Babel'' that undermines message effectiveness is acute during contexts like the COVID-19 pandemic.

Communicators being on the same page is particularly important for multifaceted and/or overlapping hazard events, as it can exacerbate the already devastating impacts of single hazard events (e.g., hurricanes, tornados etc.) by compounding effects in which external environmental conditions can lead to concurrent and cascading emergency situations. For instance, natural disasters can in some settings increase the risk of infectious disease \cite{waring_threat_2005}, via a process in which the natural disasters create disruption of critical infrastructure and displace people, providing contexts (like the close, crowded quarters of emergency shelters) that lead to environments conducive to disease spread \cite{kouadio_infectious_2012}. Disruptions to food and drinking water supplies can further exacerbate conditions and lead to other health issues \cite{connolly_communicable_2004}. Additionally, these contexts may lead to lapses in disease prevention, resulting in the potential resurgence of previously-controlled diseases that can continue the devastation beyond the original crisis-impacted location \cite{connolly_communicable_2004}.  Similar effects can occur with chronic (non-infectious) disease when vulnerable populations are unable to access needed medicines, medical devices, or services, or are simply moved to conditions in which they are unable to persist in health-maintenance activities.  Effectively responding to such cascading hazards can require complex or contingent messaging, which may be undermined by incompleteness or inconsistency.

\subsection{Information (Re)Transmission}

One important measure of successful communication strategies is a message's likelihood to be passed on to others by the initial audience members, also known as message retransmission. Message retransmission has had a rich tradition of study within sociology and public communications as a form of message engagement \cite{allport_psychology_1947, shibutani_improvised_1966}. Message retransmission is of particular interest to agencies communicating with the public, for several reasons. First, the act of message retransmission is an indicator of active engagement with a message, indicating that the message was salient to retransmitter; it can thus be used as a behavioral indicator of saliency. Second, retransmitted messages increase the likelihood of greater exposure to the public beyond the original communicators' initial audience. Repeated exposures to the same message either through repetition by an entity or through retransmitted instances of a message have further been shown to impact the likelihood of behavioral change \cite{centola_spread_2010}. Given its importance, there is unsurprisingly a great deal of literature on how the amplification of a risk-related message (in particular) may either amplify or attenuate how risks are perceived by others \cite{renn_social_1992,pidgeon_social_2003}.

Several studies in this vein have investigated messaging strategies and engagement of online crisis and public health communications \cite{sutton_cross-hazard_2015, sutton_getting_2019}, including the recent case of the COVID-19 pandemic \cite{sutton_covid-19_2020, renshaw_cutting_2021, mazid_social_2022, depaula_platform_2022}. These studies focused primarily on the factors that relate to domain-specific keywords and hashtags used in messages and the use of other digital affordances, and how these on average influence various factors of engagement (particularly retransmission). However, these studies do not consider the larger discourse the communications were embedded in, or the strategic usage of the aforementioned causal narratives, instead focusing on the particular features of the messages themselves. This research aims to build upon this existing literature within the same context to provide a deeper understanding of the broader dialogue at hand.

\subsection{COVID-19: Causal Narrative Structure and Strategy}

Given the above, our study attempts to bridge a gap between the importance of causal narratives used in the COVID-19 communication discourse -- paying particular attention to the structure of that discourse, as well as investigating the relative importance of different causal narrative strategies leveraged by communicators during this turbulent period. Given the case for message consistency across time and between communicators, this work seeks to understand just how consistent official messaging is within this context. Our study also seeks to understand just how the structure of the semantic network and themes utilized in official causal narratives impacts message retransmission which, as previously discussed, acts as a critical indicator of message reach.

\section{Data and Methods} \label{sec_methods}

For this work, we leverage a census of nearly 1 million Twitter communications ($n=989,340$) disseminated between January 2020 and mid-March 2021, by primarily U.S. based public health agencies ($n=411$), federal and state emergency management agencies ($n=50$), state governors ($n=54$), local mayors ($n=104$) and local emergency management agencies ($n=104$), with local accounts drawn from the 100 largest cities in the United States. Accounts were chosen to reflect the state of online communications during COVID-19 from public officials and organizations, with a large coverage across the U.S. population.\footnote{Further discussion about the consideration of how accounts were selected can be found in \citeNP{sutton_covid-19_2020} and \citeNP{renshaw_cutting_2021}.} We consider instances of original as well as retransmitted messages for this work, as this reflects what is transmitted to an official or organization's audience, whether it is a message of their own creation or of someone else's.

Twitter messages were collected using the Twitter representational state transfer (REST) application programming interface (API), which allowed for the collection of tweets for specific accounts, including the tweet content and associated metadata. All messages and associated attributes were publicly available for collection.

\subsection{Extraction of Causal Units}

As our focus here is on causal assertions, we examine the subset of messages containing causal units (i.e., two elements linked by a causal connective).  We identify messages by phrasal templates, with qualifying messages containing elements linked by the phrases ``due to,'' ``because of,'' or ``caused by;'' filtering for these terms led to a corpus of 11,514 messages.  The resulting messages were then parsed to obtain the templated phrases corresponding to putative causes and effects (splitting the message at the causal connective, and with assignment of cause and effect based on standard usage for the phrase in question).  For instance, the hypothetical message ``Square Toe County shelters are closed until further notice due to COVID-19,'' ``COVID-19'' is the asserted cause, and ``Square Toe County shelters are closed until further notice'' is the asserted effect.  We refer to these elements of each message as the \emph{causal subpart} and the \emph{effect subpart,} respectively.  Due to the ambiguity of phrasal parsing, instances in which the causal connective occurred at the beginning of a sentence were excluded from analysis.

\subsection{Lexicon}
\label{sec:lex}

While extracted message components occasionally consist of simple and obvious terms or phrases (e.g., ``COVID-19''), they are typically more elaborate.  For instance, sentences such as ``[We ask you to practice social distancing] due to [COVID-19]'' were typical (parsing brackets added), in which at least one of the causal subpart or the effect subpart were complex phrases or clauses (here, the effect subpart ``We ask you to practice social distancing'').  Such complex subparts were mapped to concepts using a lexicon.  The lexicon consisted of keywords and phrases based on prior work within the hazard communication literature, particularly relating to disaster communications, while accounting for language and content observed during the COVID-19 pandemic. Lexicons have been successfully used for message classification in prior work on hazards \cite{imran_practical_2013,olteanu_crisislex_2014}, and have been employed successfully on social media communications by public agencies during COVID-19, in particular \cite{sutton_covid-19_2020, renshaw_cutting_2021}.  Our lexicon is based on these latter examples.  We note in passing that lexicon-based concept coding works well in settings in which speakers use fairly standardized language, on a relatively focused set of concepts, and do so consistently throughout the corpus.  Although much informal communication does not satisfy these conditions, they are frequently met by public-facing remarks by public agencies, particularly in hazard settings (and particularly when subject to the constraints of terse messaging).  This strategy thus works well here, but may not be effective in studies using data from other sources.

All work with identifying lexical categories was performed by a manual coder. Lexical categories were created by reviewing the two corpuses of extracted materials: the subparts containing the causes and the subparts containing the effects. For each corpus, a random sample of subparts were manually reviewed to identify keywords that completely characterize the entire subpart. For instance, in the effect subpart ``We ask you to practice social distancing'' the most relevant keywords to the concept are ``social distancing.'' The identified keyword is then searched for using regular expressions in the entire corpus of effect subparts. If all returned subparts were successfully characterized by the keyword, then all subparts were categorized as such and removed from further consideration. If only a proportion of subparts were successfully characterized by keyword, then they were characterized as such, removed from further consideration, and the rest are returned to the consideration pool. Similar keywords identified by the coder were tried first before returning to random sample pulls for further keyword identification. This process was repeated until subparts left in the corpus for consideration were not easily identifiable through keywords, and were withdrawn from further analysis.

These keywords were then manually reviewed and gathered into broader keyword classes, characterized by concepts of the content itself. These concepts were based on a previous lexicon of Sutton et al, (\citeyear{sutton_covid-19_2020}), developed for this particular use case. For example, the keywords ``social distancing,'' ``stay home,'' and ``mask'' were classified under the broader concept of actions and efficacy measures taken against COVID-19 (``Actions/Efficacy"). All subparts were then classified into concepts based on the identified keyword within. For instance, all subparts containing the keyword ``social distancing'' were classified as pertaining to ``Actions/Efficacy.'' For the purposes of retransmission analysis and visualization, concepts were then grouped based on broader themes of messaging. For instance, the concepts ``Actions/Efficacy", ``Vaccination", and ``Restrictions" all speak of measures the public can take to safeguard against disease and thus are classified as pertaining to the larger theme of ``Primary Threat Measures".  

\subsection{Network Formulation and Analysis}

As our interest is in the structure of the full discourse of causal claims made by the actors in our sample, we aggregate individual causal units extracted from individual messages into (causal) semantic networks.  Specifically, we define a valued digraph on the set of all identified concepts, in which the value of the $A \to B$ edge is the number of observed assertions in which a subpart mapped to concept $A$ was alleged to cause a subpart mapped to concept $B$.  The valued network then represents the overall discourse of causal claims, weighted by the frequency with which they were invoked; dichotomization of this network leads to a digraph containing all causal linkages invoked during the period.  We here examine the total aggregate network based on all messages sent during the period, as well as the networks formed by stratifying messages by time or by account.  Time-varying networks were formed by binning messages by month of initial posting, leading to 15 monthly networks (January 2020 through March, 2021).  Account-stratified networks were formed by binning messages respectively posted by public health agencies, state/federal emergency management agencies, and elected officials (gubernatorial and mayoral accounts).  Comparison of these networks allows us to examine how discourse varies over time and by the type of agency involved.

We examine the structure of causal claims within public agency discourse by analysis of the above semantic networks.  Previous research on semantic networks has suggested that concepts with many direct relations are more salient than concepts with few direct relations \cite{trabasso_causal_1982,trabasso_causal_1985,danowski_organizational_1988,graesser_prose_2013} suggesting that messages containing assertions involving high-degree concepts may have enhanced salience (and thus greater likelihood of being passed on to others).  As noted earlier, consistency is a well-known desiderata of public communication, but one that is difficult to achieve via decentralized communications; we thus also consider the extent that similar patterns of claims across types of actors and over time.  Put differently, we investigate whether the causal narratives used in public agency communications reflected a shared, \emph{consensus} discourse, or whether narratives articulated competing points of view.

Consensus in semantic networks can be gauged by the extent to which concept relations are shared amongst actors and how strongly they are shared \cite{carley_semantic_1993}. To examine the shared structural features of the five different role-specific networks, we perform a network principal component analysis (PCA) on the valued networks, in which an eigendecomposition of the role graphs yields a representation of the semantic networks as linear combinations of orthogonal valued graphs. In other words, for a graph covariance matrix $C$ for $p$ graphs $G_1, ..., G_p$, the eigendecomposition of $C$ is $C=W\Lambda W^T$. The eigenvectors $W$ yield the loading of each graph on each principal component. The score graph for each principal component is then a valued network representing a particular dimension or facet in terms of which the observed networks can be expressed.  As we show, the sets of networks studied here have negligible weight beyond the first two components, and our analysis thus focuses on the loadings and score graphs of the first two principal components. Interpretively, the first principal component yields a network representation of the causal relationships that tend to be held in common throughout all the networks being analyzed\footnote{This interpretation holds when the graph covariance matrix has a principal eigenvector, as is always the case in our analyses.} - the dominant \emph{underlying theme} or central tendency of the set as a whole. The second principal component represents the strongest \emph{contrast} in causal assertions among the semantic networks, identifying the assertions that most strongly distinguish the respective semantic networks from each other.  Likewise, the corresponding eigenvalues intuitively reflect the relative weight or importance of these two components: if the discourse is primarily consensual, the first eigenvalue will be substantially greater than the second, while comparable values suggest a high degree of disagreement between articulated views.  A more detailed description of network PCA may be found in \citeNP{butts_social_2008}.

As previously discussed, consistency across time is also critical for building credibility and behavioral uptake of risk-minimizing behaviors. However, during the observed time period of our dataset, officials and organizations gained a greater understanding of the SARS COV-2 virus along with consequences of a global pandemic. To examine these changes, we thus examine the incidence of commonly-employed causes and effects in causal narratives as the pandemic developed, and consider how communications reflected changes in the pandemic and events revolving around secondary threats.

\subsection{Predicting Retransmission}
\label{sec:predretran}

As noted above, the position of a causal claim within the discourse is hypothesized to relate to its saliency, and to engagement with the associated message.  To examine the relationship between engagement and semantic network structure, we perform a negative binomial regression of the number of times a message containing a causal unit is retransmitted on properties of the position of that unit and its elements within the discourse network. Message retransmission is a widely studied form of message engagement, serving to amplify the original message. Message retransmission in a hazards setting has been recognized to be of critical importance since long before the era of digital communications \cite{danzig_effects_1958,allport_psychology_1947}, and recent work has continued this line of inquiry in the social media context \cite{sutton_cross-hazard_2015,sutton_getting_2019, olson_build_2019} (including in the context of the COVID-19 pandemic \cite{sutton_covid-19_2020,renshaw_cutting_2021}).  Here, we are primarily interested in the impact of discourse position on message transmission.  Specifically, we hypothesize:

\begin{itemize}
\item[] H1: \emph{Ceteris paribus}, a message is more likely to be retransmitted if the attributed effect has many consequences. I.e., given a cause and effect pair $i,j$, if there are many consequences to $j$ (high outdegree), then messages containing an $i,j$ causal unit will be retransmitted at a higher rate.
\item[] H2: \emph{Ceteris paribus}, a message is more likely to be retransmitted if the cause itself has many causes. I.e., given a cause and effect pair $i,j$, if $i$ has a high indegree, then messages containing an $i,j$ causal unit will be retransmitted at a higher rate.
\item[] H3: \emph{Ceteris paribus}, a message is less likely to be retransmitted if the contained causal Unit exhibits transitive closure with another concept as a mediator - if there are other ways within the discourse to form a narrative in which $i$ leads to $j$, then messages asserting an $i,j$ causal unit are redundant and less likely to be passed to others.
\item[] H4: \emph{Ceteris paribus,} message retransmission will vary depending on the themes with which its causal elements are associated.
\item[] H5: \emph{Ceteris paribus}, as a cause is used more in discourse, messages containing the cause are less likely to be retransmitted. In other words, there is a topic ``fatigue,'' with increasingly familiar causes becoming less salient with repeated exposure.
\item[] H6: Similar to H5, as an effect is used more in discourse, messages containing the effect are \emph{ceteris paribus} less likely to be retransmitted.
\end{itemize}

We assess these hypotheses via a message-level regression model, taking the number of times each message containing a causal unit was retransmitted as the dependent variable.  Predictors include the number of attributed effects of the cause and effect pair found within the message, the number of precursory causes to a cause and effect pair, whether the causal relation within the message is the transitive closure of a two-path involving a third concept, indicators for the broader themes to which the cause and effect respectively belong, and frequency of cause and effect usage up until the month-year of the message's creation. The structural variables are operationalized from the conceptual, dichotomous discourse network whereas the fatigue-terms are operationlized from the valued discourse network. As is typical for Twitter data, many posts are not retweeted (22\% of tweets in this dataset) while few were extremely well-retweeted (max=199,760). Due to the extreme right-skew, we conduct our regression using the negative binomial family which accounts for unobserved hereogeneity in our retransmission counts. This follows prior literature on modeling overdispersed count data in the domain of official COVID-19 communications \cite{sutton_covid-19_2020,renshaw_cutting_2021}. All regression analysis was conducted in R using the \texttt{glmmADMB} package \cite{r_core_team_r_2021,fournier_ad_2012}.

\section{Results} \label{sec_results}
\subsection{Cause and Effect Extraction}
Using the lexicon discussed in Section \ref{sec:lex}, approximately 390 keywords and variants were discovered for the corpus of causes and approximately 260 were discovered for the corpus of effects. Classification of these keywords yield the discovery of 39 different concepts and, of the 15,949 tweets containing one of the selected causal phrases, 11,514 messsages were successfully coded as relations between these concepts. Broadly, the concepts encompass the Primary Threat of COVID and related attributes such as Susceptibility and the Severity and Impact of the disease, the far-reaching consequences of COVID such as in restrictions and financial impacts, Secondary Threats such as weather and illnesses or injury, Events and Actors mentioned, and expressive sentiments such as resilience and gratitude. These 39 concepts are used in the creation of the causal narrative networks, in which nodes are concepts that may be used as causes and/or effects. Further classification of the 39 concepts into broader content themes yielded thirteen themes, which are primarily used to simplify the engagement analysis. Descriptions of the concepts with message examples classified by the content themes are presented in Table \ref{tab:top_desc}.

\begin{singlespace}

	\begin{longtable}{L{2cm}|L{3cm}|L{5cm}|L{5cm}} \hline\hline
			 \textbf{Theme} & \textbf{Concept}                         & \textbf{Definition}                                                                         & \textbf{Example}                                                                             \\
   \hline
   			\textbf{Primary Threat} & \textbf{Primary Threat}    & Directly describes COVID-19                                                                 & ``\#COVID19.", ``the coronavirus pandemic." \\
    \hline
			\textbf{Susceptib-ility} & \textbf{Susceptibility}            & Describing individuals or groups at risk of COVID-19                                        & ``the increased risk of \#COVID 19 spread onboard ships" \\
   \hline
			\textbf{Primary Threat Impact} & \textbf{Severity/Impact}           & Relating to the magnitude or extent of the impact of COVID-19                               & ``having COVID-19 case rates three times higher than the state's numbers"                     \\
   		& \textbf{Testing}                   & Relating to COVID-19 testing availability and procedures                                    & ``limitations in testing opportunities."                                                      \\
      	& \textbf{Deaths}                    & Relating to deaths resulting from COVID-19                                                  & ``Today, nearly one year after we mourned the first death"                                    \\
			& \textbf{Losses}                    & Relating to the loss of lives or resources                                                  & ``There are a lot of people who are losing their lives"                                       \\
   \hline
			\textbf{Primary Threat Measures} & \textbf{Actions/Efficacy}          & Recommended protective measures against COVID-19                                            & ``the \#COVID 19 response"                                                                    \\
			& \textbf{Vaccination}               & Relating to vaccinations against COVID-19                                                   & ``limited vaccine supply."                                                                    \\
			& \textbf{Travel}                    & Relating to travel                                                                          & ``Travel will be very difficult"                                                              \\
			& \textbf{Restrictions}              & State-mandated restrictions to normal operations                                            & ``restrictions enacted to fight Covid-19"                                                     \\
			& \textbf{Isolate}                   & Quarantine and isolation measures against COVID-19                                          & ``informal gatherings."                                                                       \\
	\hline		
            \textbf{Spread} & \textbf{Spread}                    & Relating to the transmission of disease                                                     & ``known exposure to a positive case"                                                          \\
            \hline
			\textbf{Economic and Financial Support} & \textbf{Economic Impacts}          & Relating to broader economic impacts to businesses and the greater economy                  & ``We are still reeling from an economic crisis"                                               \\
			& \textbf{Financial Struggle}        & Relating to personal financial struggle                                                     & ``If you have experienced financial hardship"                                                 \\
			& \textbf{Need Assistance}           & Identifying individuals and groups in need of assistance                                    & ``Do you need help paying your rent"                                                          \\
			& \textbf{Provide Assistance}        & Source is providing resources as assistance                                                 & ``Texas update: additional counties and funds have been approved for disaster relief efforts" \\
			\textbf{Data Processing} & \textbf{Data}                      & Relating to data management                                                                 & ``a backlog of over 2,000 results received from Thursday through Sunday."                     \\
   \hline
			\textbf{Transitions and Shifts} & \textbf{Disruptions} & Changes in times for regularly scheduled operations, including cancellations and delays     & ``The County Courthouse remains closed to the public"                                         \\
			& \textbf{Change of Mode}            & Changes in format for regularly scheduled operations, including switches to virtual formats & ``This year, we met virtually"                                                                \\
			\hline
            \textbf{Official Response} & \textbf{Official Response}         & Official mandates against COVID-19                                                          & ``One year ago today, I declared a state of emergency"                                        \\
			\hline
			\textbf{Secondary Threats} & \textbf{Mental}                    & Relating to mental disorders                                                                &``Coping with the stress"                                                                     \\
			& \textbf{Food}                      & Relating to food, including food insecurity, recalls, etc.                                  & ``In need of food"                                                                            \\
			& \textbf{Blood}                     & Relating to state blood supplies                                                            & ``There continues to be a shortage of blood"                                                  \\
			& \textbf{Weather}           & Relating to weather conditions                                                  & ``incoming lightning and severe weather."                                                     \\
			& \textbf{Infrastructure}      & Relating to local infrastructure such as damages and power outages                                                & ``Are you facing costly damages"                                                            \\
			& \textbf{Preparedness}              & Recommendations for preparing against secondary threats                                     & ``Be sure to have an emergency plan if you need to take immediate action"                     \\
			& \textbf{Traffic}                   & Relating to traffic, roads, etc.                                                            & ``a serious collision."                                                                       \\
			& \textbf{Illness/Injury}            & Relating to illnesses outside of COVID or injuries                                          & ``a broken bone or severe injury is a risk factor for blood clots."                           \\
			& \textbf{Non-COVID Deaths}          & Deaths resulting from non-COVID related reasons                                             & ``a non-COVID related death determination."                                                   \\
			& \textbf{Other Secondary Threats}   & Relating to other secondary matters such as state infrastructure and violence               & ``a sewage spill."                                                                            \\
			\hline
			\textbf{Emotional Responses and Coping} & \textbf{Gratitude}                 & Expressions of thanks                                                                       & ``the generosity of our supporters."                                                          \\
			& \textbf{Resilience}                & Expressions of community resilience                                                         & ``Our state is a better place"                                                                \\
			& \textbf{Challenges}   & Expressions of difficulties faced by the general public or certain groups                   & ``Over the last year, the restaurant industry faced unprecedented challenges"                 \\
			\hline
			\textbf{Events and Actors} & \textbf{Demographics}              & Describing the demographics of a particular group                                           & ``their sexual orientation and/or gender identity."                                           \\
			& \textbf{``You"}                     & Direct indication of the user viewing the message                                           & ``you, your work and your compassion "                                                        \\
			& \textbf{Other Actors}              & Direct indication of actors or groups other than the user viewing the message               & ``him."                                                                                       \\
			& \textbf{Events}                    & Specific events and holidays                                                                & ``protests in Downtown ATL."                                                                  \\
			\hline
			\textbf{Off-Topic} & \textbf{Off-Topic}                 & Topics not relating to COVID-19 or secondary threats                                        & ``improper firework disposal" \\ \hline\hline
   
   \caption{Concepts obtained from manual coding process on messages with selected causal phrases, complete with concept definition and an accompanying example. Concepts have been grouped into broader themes, which are used in retransmission analysis. \label{tab:top_desc}}
	\end{longtable}

\end{singlespace}

\subsection{Overall Network}

To describe the conceptual linkages between concepts, we focus on describing the dichotomized network of linkages between causes and effects. The overall conceptual network of causal narratives is weakly connected with a density of 0.26 (mean in/outdegree 10.14).  We thus find that, on average, each concept within the discourse is linked to a fairly large number of other concepts, implying a fairly complex causal narrative. Contrary to the intuition of a unidirectional (cause $\to$ effect) discourse, we see a fairly high degree of reciprocity, with the probability of an $A \to B$ assertion given a $B \to A$ assertion being approximately 40\% higher than the marginal probability of an $A \to B$ assertion.  This implies a high degree of reciprocal causation, at least at the concept level.  The network also exhibits a relatively high level of indegree, outdegree, and betweenness centralization measure, suggesting that a small proportion of concepts have much higher numbers of causal associations than average, and likewise that a small number of concepts lie on disproportionately many deductive pathways (i.e., $A \to B \to C$); conditional uniform graph (CUG) tests for marginal network properties are shown in Table~\ref{tab:cug_results}.  Many concepts have a non-zero net degree (outdegree - indegree), suggesting an overall tendency for concepts to be more often framed as causes or as effects (though most concepts play both roles). See Table \ref{table:netdeg} for more information on out-degree, in-degree, and net degree.

\begin{table}[h]
	\centering
	\begin{tabular}{ll|ccc} \hline\hline
		Statistic         & Conditioned On & Obs. Value & $\Pr(X\geq \mathrm{Obs})$ & $\Pr(X\leq \mathrm{Obs})$ \\
		\hline
		Edgewise Reciprocity            & Edges                    & 0.368               & $<0.001$                                & $>0.999$                            \\
		Transitivity               & Dyad Census              & 0.582               & $<0.001$                                & $>0.999$                             \\
		In-Degree Centralization   & Dyad Census              & 0.538               & $<0.001$                                & $>0.999$                             \\
		Out-Degree Centralization  & Dyad Census              & 0.512               & $<0.001$                                & $>0.999$                             \\
		Betweenness Centralization & Dyad Census              & 0.086               & $<0.001$                                & $>0.999$                        \\    \hline \hline
	\end{tabular}
	\caption{Conditional uniform graph tests for the combined dichotomized discourse network. \label{tab:cug_results}}
\end{table}

\newpage

\begin{singlespace}
\begin{table}[H]
\centering
\begin{tabular}{lrrr}
  \hline\hline
Concept & Out-Degree & In-Degree & Net Degree (Out-In) \\ 
  \hline
Primary Threat & 4293.00 & 66.00 & 4227.00 \\ 
  Disruptions & 216.00 & 3503.00 & -3287.00 \\ 
  Weather & 2680.00 & 134.00 & 2546.00 \\ 
  Deaths & 0.00 & 773.00 & -773.00 \\ 
  Financial Struggle & 99.00 & 742.00 & -643.00 \\ 
  Need Assistance & 0.00 & 611.00 & -611.00 \\ 
  Actions/Efficacy & 55.00 & 573.00 & -518.00 \\ 
  Illness/Injury & 576.00 & 172.00 & 404.00 \\ 
  Mental & 10.00 & 408.00 & -398.00 \\ 
  Official Response & 18.00 & 382.00 & -364.00 \\ 
  Severity/Impact & 378.00 & 741.00 & -363.00 \\ 
  Infrastructure & 168.00 & 511.00 & -343.00 \\ 
  Provide Assistance & 3.00 & 191.00 & -188.00 \\ 
  Other Secondary Threats & 330.00 & 160.00 & 170.00 \\ 
  Traffic & 323.00 & 154.00 & 169.00 \\ 
  Events & 166.00 & 0.00 & 166.00 \\ 
  Travel & 5.00 & 156.00 & -151.00 \\ 
  Losses & 13.00 & 159.00 & -146.00 \\ 
  Economics & 122.00 & 238.00 & -116.00 \\ 
  Data & 100.00 & 0.00 & 100.00 \\ 
  Spread & 114.00 & 30.00 & 84.00 \\ 
  Demographics & 84.00 & 0.00 & 84.00 \\ 
  Isolate & 34.00 & 111.00 & -77.00 \\ 
  Susceptibility & 37.00 & 111.00 & -74.00 \\ 
  Food & 16.00 & 88.00 & -72.00 \\ 
  Testing & 72.00 & 0.00 & 72.00 \\ 
  Blood & 3.00 & 73.00 & -70.00 \\ 
  Drugs & 75.00 & 9.00 & 66.00 \\ 
  Vaccination & 64.00 & 0.00 & 64.00 \\ 
  Resilience & 9.00 & 64.00 & -55.00 \\ 
  Gratitude & 42.00 & 0.00 & 42.00 \\ 
  Restrictions & 40.00 & 5.00 & 35.00 \\ 
  Change of Mode & 2.00 & 34.00 & -32.00 \\ 
  Non-COVID Deaths & 24.00 & 0.00 & 24.00 \\ 
  Challenges & 181.00 & 157.00 & 24.00 \\ 
  You & 23.00 & 0.00 & 23.00 \\ 
  Off-topic & 9.00 & 29.00 & -20.00 \\ 
  Other Actors & 14.00 & 0.00 & 14.00 \\ 
  Preparedness & 0.00 & 13.00 & -13.00 \\ 
   \hline\hline
\end{tabular}
\caption{Outdegree, indegree, and net degree of each of the 39 concepts employed in observed causal units. Out-degree signifies the number of times the concept was mentioned as a cause, whereas in-degree signifies the number of time said concept was mentioned as an effect. Table is sorted in magnitude of net degree to highlight concepts that are disproportionately employed either in cause or effect roles. \label{table:netdeg}}
\end{table}
\end{singlespace}

\begin{figure}[H]
\centering
\includegraphics[width=.8\textwidth]{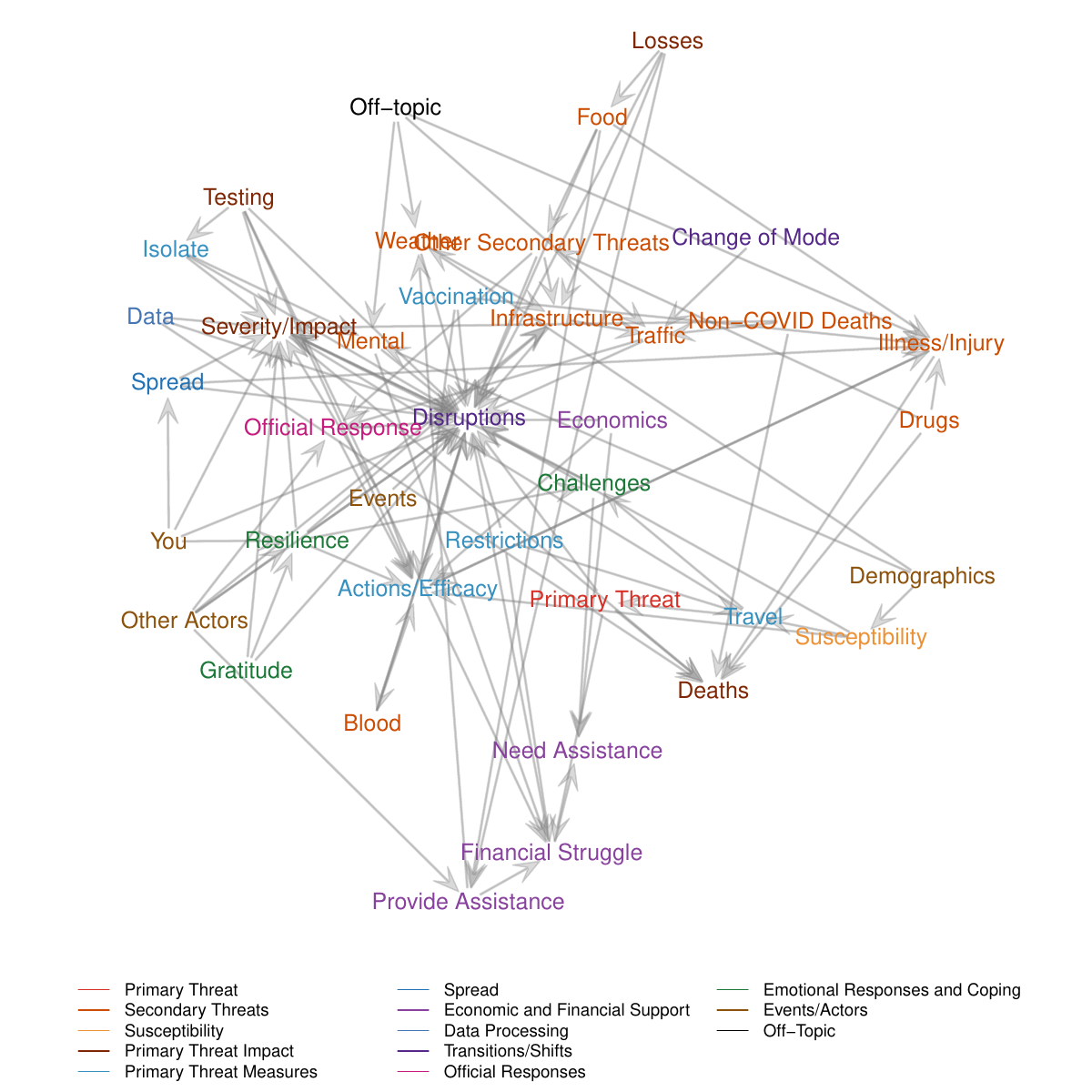}
\caption{Overall causal narrative network, only showing edges between three strongest effects for each concept to highlight well-utilized causes.}
\label{fig:net_effect}
\end{figure}

\begin{figure}[H]
\centering
\includegraphics[width=.8\textwidth]{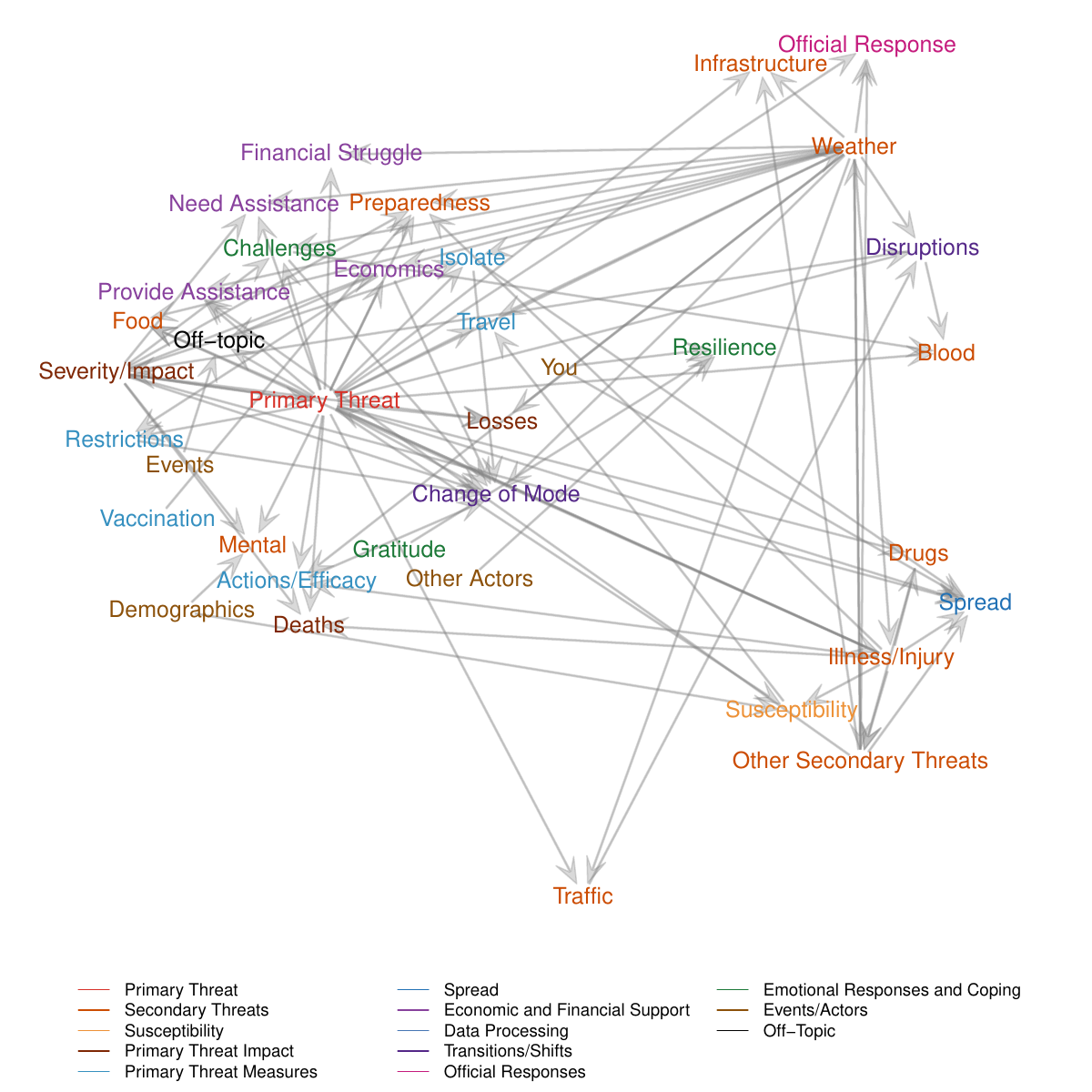}
\caption{Overall causal narrative network, only showing edges between three strongest causes for each concept to highlight well-utilized effects.}
\label{fig:net_cause}
\end{figure}

The most-utilized causal concept was ``Primary Threat,'' i.e., direct mentions of the COVID-19, comprising 38\% of the utilized causes, followed by ``Weather'' (28\%) and ``Illness/Injury'' (6\%). This suggests that the most salient threats of this time period with the largest number of consequences were directly from COVID and weather conditions, with the other classes of causes including complications of COVID and Secondary Threats taking a sideline. See Figure \ref{fig:net_effect} for a visualization of the causal narrative network, in which only the three strongest effects are shown for each concept, thereby highlighting which concepts are often referred to as an effect. By far, the most-utilized effect class was ``Disruptions,'' or disruptions to operations, comprising 31\% of the effects utilized. This is in alignment with previous literature in that many communications from these organizations and officials pertain to changes in scheduled services \cite{sutton_cross-hazard_2015, sutton_covid-19_2020,renshaw_cutting_2021}. See Figure \ref{fig:net_cause} for a visualization of the causal narrative network, in which only the three strongest causes for each concept are highlighted.  

As expected, the most-utilized causal narratives (e.g. cause and effect pairs) were often pairings of the most-utilized causes and effects, with ``Weather'' resulting in ``Disruptions'' (13.3\% of narratives used), ``Primary Threat'' resulting in ``Disruptions''  (8.3\%),  ``Primary Threat'' resulting in ``Financial Struggle''  (5\%), ``Weather'' resulting in other ``Weather'' effects (4.5\%) and ``Primary Threat'' resulting in ``Deaths'' (4.2\%). See Table \ref{tab:narr_ex} for examples of each of these common narratives. Note that the discrepancy between weather-related disruptions and primary threat-related disruptions is not simply due the first two months of the dataset taking place in the winter before the national pandemic lockdown, with only 1\% of weather-related disruptions being communicated before March of 2020. These results underline how the usual discourse of such organizations and officials were altered by the COVID-19 pandemic, with the effects of the primary threat of COVID-19 becoming a major focus of public-facing discourse. Although many disruptions of scheduled operations were impacted by COVID, weather effects continued to have considerable impacts throughout the pandemic, as well.

\begin{table}[H]
\centering
\begin{tabular}{ll|p{9cm}} \hline\hline
    Cause          & Effect             & Example                                                                                                   \\
    \hline
    Weather        & Disruptions        & ``UPDATE: The testing site at Sunnyview will be CLOSED on Thursday, January 28 \textit{due to} weather conditions." \\
    \hline
    Primary Threat & Disruptions        & ``Check your voting location + requirements in advance because they may have changed \textit{due to} \#COVID19."    \\
    \hline
    Primary Threat & Financial Struggle & ``Help is available for those struggling to pay rent or utilities \textit{due to} COVID-19."                        \\
    \hline
    Weather        & Weather            & ``Parts of the central Gulf coast might experience flash flooding \textit{due to} heavy rains."                     \\
    \hline
    Primary Threat & Deaths             & ``There are 18,114 deaths in the State of Ohio* and 17,992 Ohio resident deaths* \textit{due to} COVID-19."      \\ \hline\hline  
\end{tabular}
\caption{Message examples of the five most used causal units. Causal phrases are italicized.}
\label{tab:narr_ex}
\end{table}

\subsection{Networks by Account Type}
\begin{figure}[H]
\centering
\includegraphics[width=.8\textwidth]{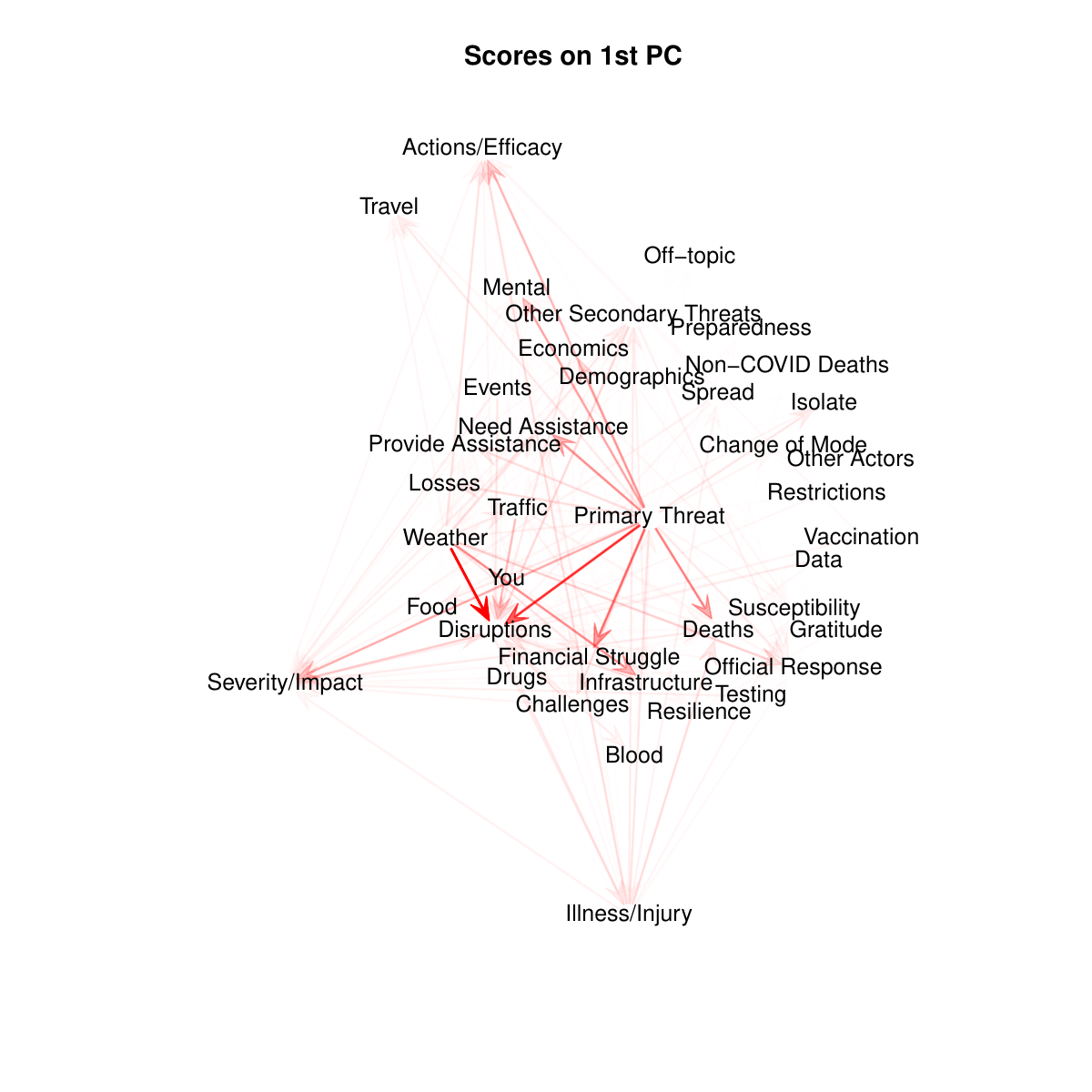}
\caption{Score graph on the first principal component.}
\label{fig:score_pc1}
\end{figure}

\begin{figure}[H]
\centering
\includegraphics[width=.8\textwidth]{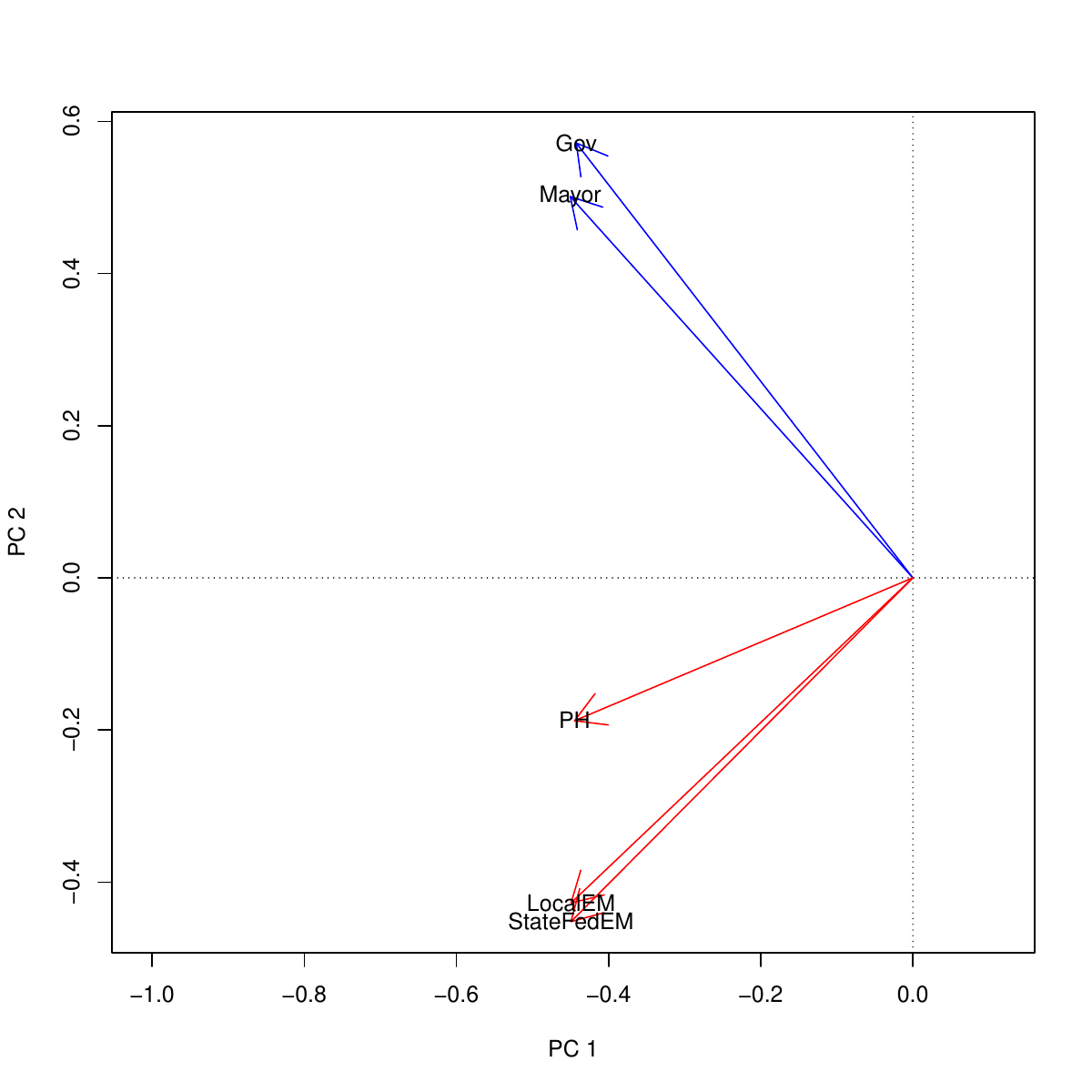}
\caption{Loadings of the five role graphs on the first two principal components of the network principal component analysis.}
\label{fig:loadings}
\end{figure}

The loadings on the first principal component are strikingly similar for all five role graphs, as can be seen from Figure \ref{fig:loadings}. From this, it is found that all account types have approximately similar weightings on the score graph of the first principal component (Figure \ref{fig:score_pc1}). As the first component of this set of networks is a principal eigenvector, the corresponding score graph represents the most commonly shared (co-moving) causal assertions seen across all five role graphs. In other words, it represents the common discourse used amongst all actors within this context. The strongest narratives used by all actors are ones in which the weather or the primary threat of COVID cause disruptions to regularly scheduled events. All actors discuss the many consequences of the primary threat, most notably on individual financial struggles and deaths.

\begin{figure}[H]
\centering
\includegraphics[width=.8\textwidth]{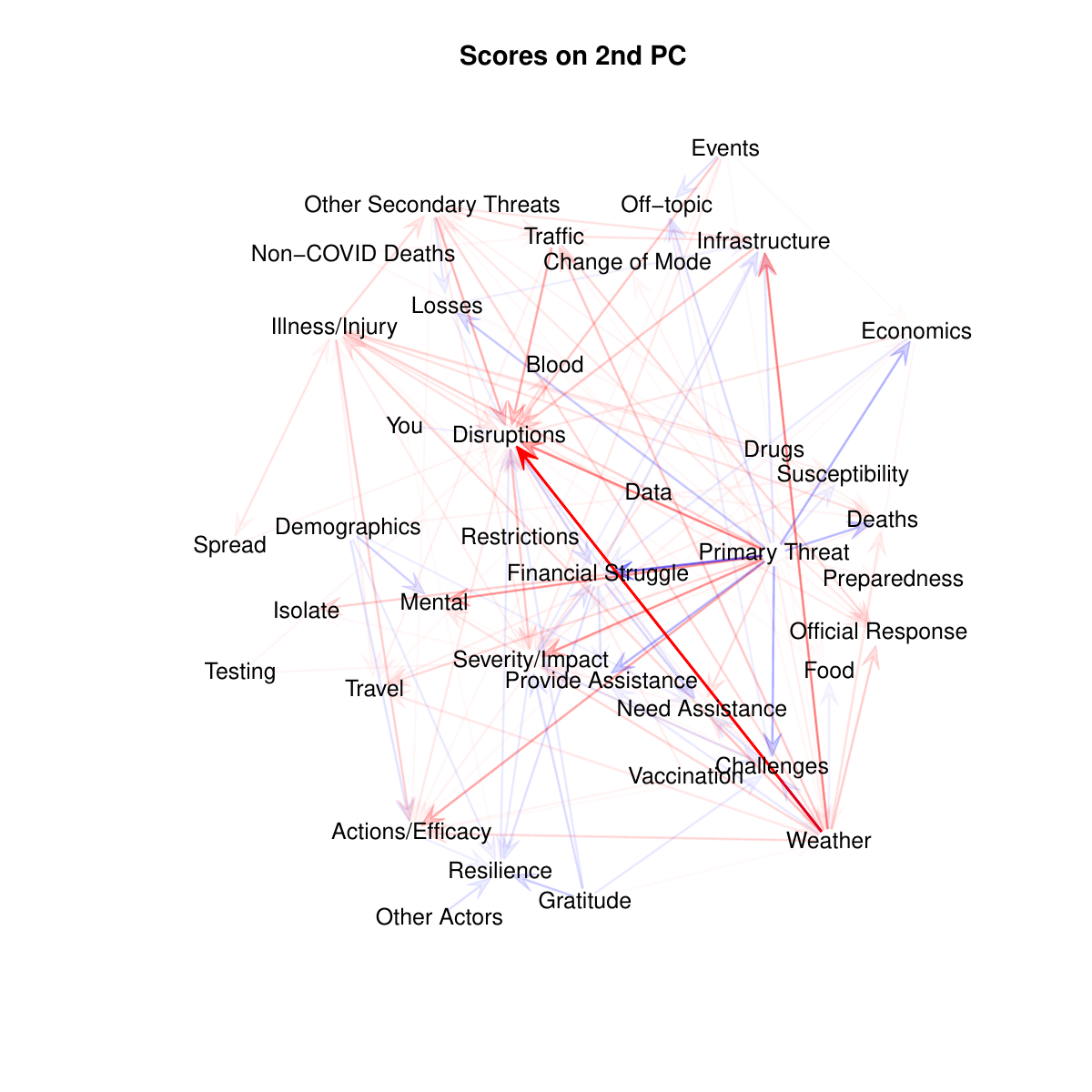}
\caption{Score graph on the second principal component.}
\label{fig:score_pc2}
\end{figure}

\begin{figure}[H]
\centering
\includegraphics[width=.8\textwidth]{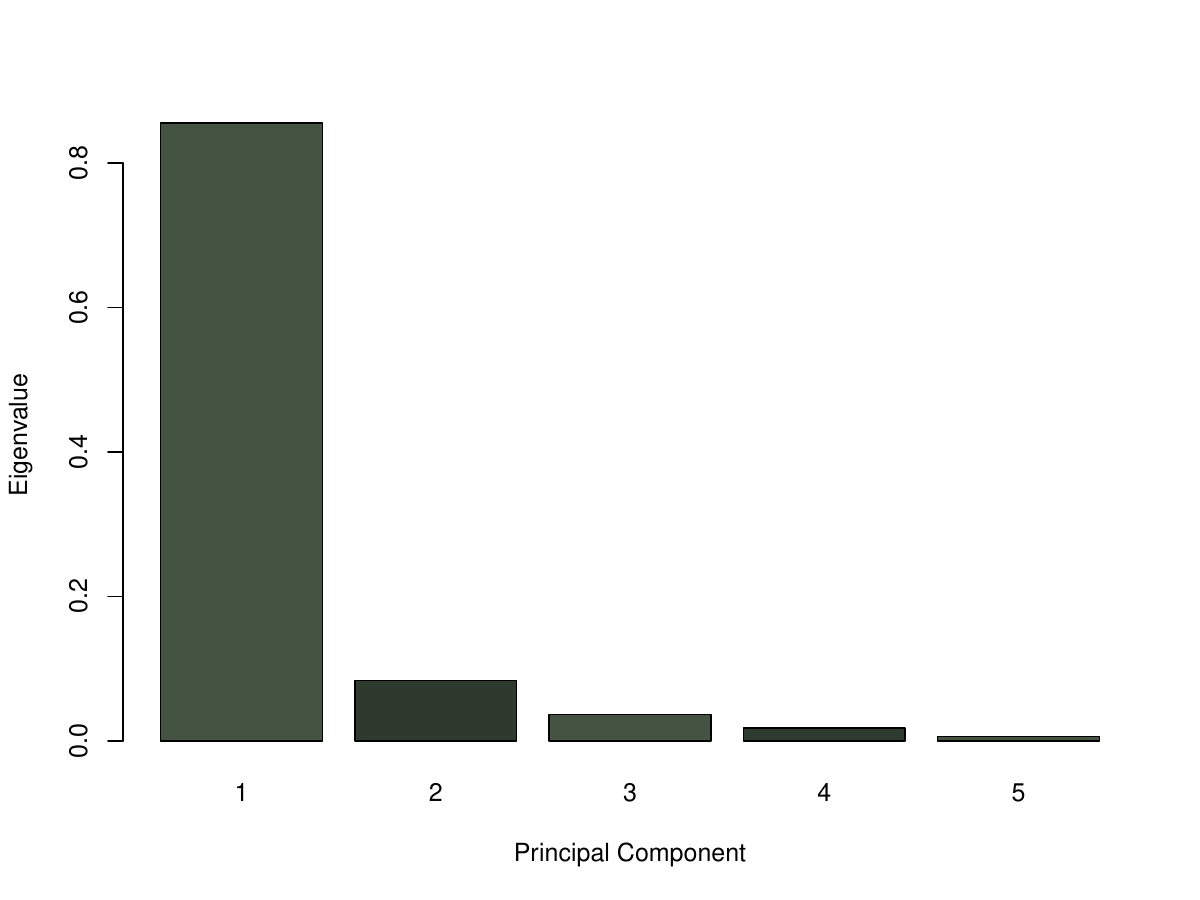}
\caption{Plot of the first five eigenvalues from the network PCA.}
\label{fig:skree}
\end{figure}

The second principal component is a contrast, splitting the actors into two different groups based on differences in their causal narrative networks. As seen from the graph loadings on the second principal component from Figure \ref{fig:loadings}, the electeds are more similar in their causal narrative usage than they are to the responding agencies. The local and state/federal emergency management agencies have similar loadings on the second principal component, whereas public health agencies have a weaker, but still negative loading. The score graph on the second principal component highlights the distinctions between the electeds and the responding agencies in their causal narrative usage (Figure \ref{fig:score_pc2}). The responding agencies are much more likely to discuss how weather conditions lead to disruptions than the electeds are. Another key difference is in the attributions made from direct mentions of the primary threat. Responding agencies are more likely to discuss the impacts of COVID-19 on the severity of cases, on disruptions, and on mental health while advising on protective actions. Elected officials, compared to the responding agencies, are more likely to discuss the impacts of the primary threat on the macro- and microeconomic situation of their constituents. Note, however, that the first eigenvalue is very large compared to the second eigenvalue (see Figure \ref{fig:skree} for a plot of the first five eigenvalues), indicating that the score graph on the first principal component contributes much more to the structure of the network set; thus, the discourse across all agencies is strongly consensual, with structural differences picked up by the second component being a relatively minor (if insightful) contributor.  We also see that components beyond the second have negligible contribution to the set of role-based discourse networks, and we thus do not pursue them further here.

\subsection{Networks by Time}

\begin{figure}[H]
\centering
\includegraphics[width=.8\textwidth]{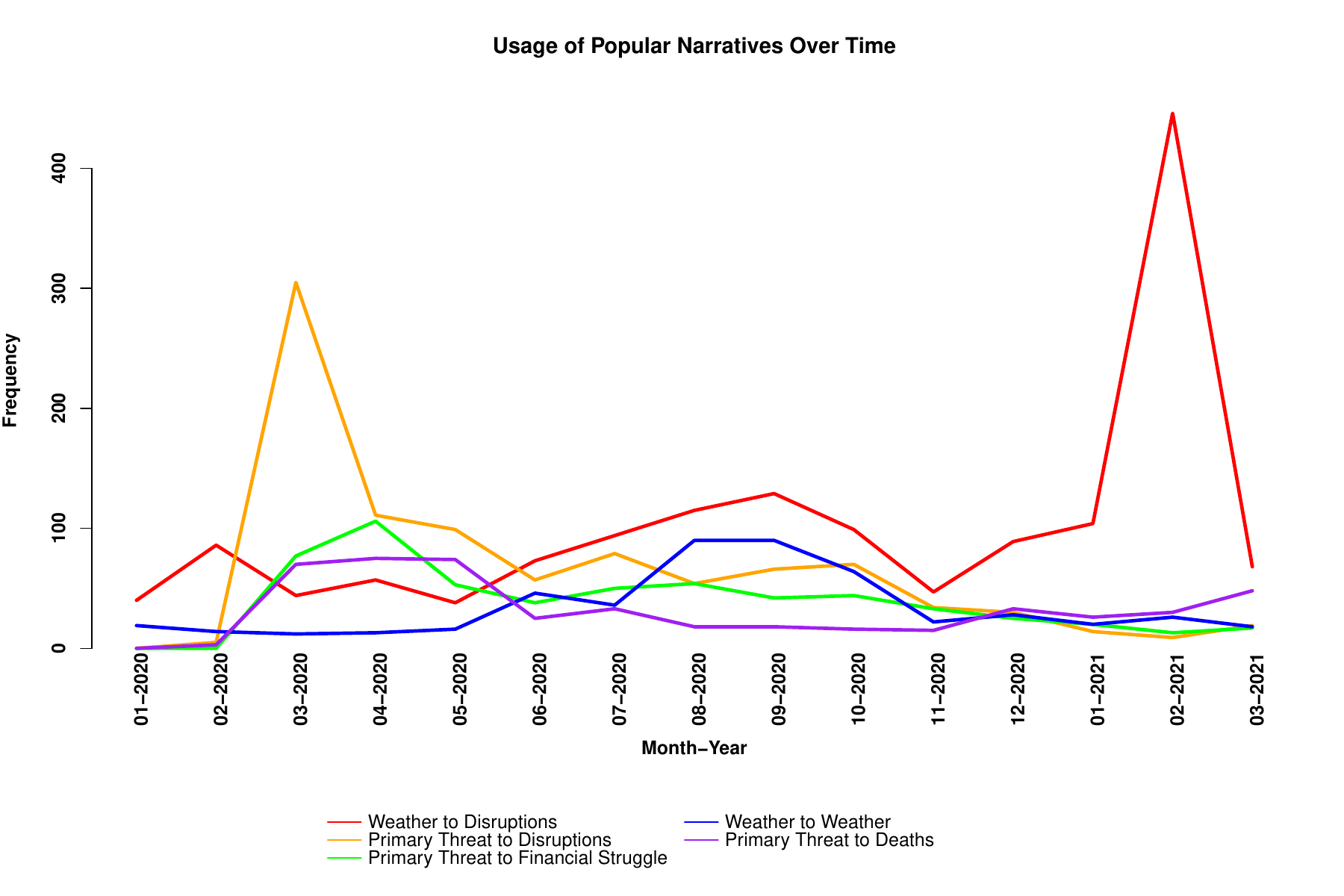}
\caption{Usage of five most utilized narratives from January 2020 to March 2021.}
\label{fig:pop_narr}
\end{figure}

Narratives are seen to change over time, with the usage of the most utilized cause and effect relations during the entire time period shown in Figure \ref{fig:pop_narr}. Before March 2020, when COVID dominated public attention, officials were observed to communicate primarily on weather-related disruptions and illnesses or injuries. After the first two months, weather-related disruptions persisted as a popular narrative while talk of illnesses and injuries took to the wayside for the primary threat.

The most popular narrative of weather-caused disruptions is seen to increase in usage over the summer natural disaster season and exhibits the largest spike in February 2021. This drastic increase is due to inclement cold weather events such as unexpectedly freezing temperatures in the state of Texas leading to hazardous conditions including power outages, water pipes bursting, and icy roadways in areas unequipped for such weather conditions \cite{noauthor_great_2023}.

In the first two months prior to the national lockdown in March 2020, much of the discourse by officials involved informing on weather-related disruptions and on illnesses/injuries. However, the conversation quickly changed to revolve around COVID-19. Discussions of the COVID-related disruptions was at its highest in March 2020, when COVID-19 took the national stage in the United States and prompted a country-wide quarantine measure. Messages with this causal narrative discussed disruptions to various aspects of daily life such as schools, medical operations, and city services. Although gradually decreasing over time, slight increases in usage in the months of July and October include warnings of gatherings for holidays such as Independence Day (July 4th) and Halloween (October 31st). 

Following nation-wide closures, there was a marked increased in discussions of individual financial impacts due to COVID-19 in April 2020, as quarantine measures were extended and employment halted for many. Messages during this time called for those facing financial struggles from the pandemic to apply to assistance programs. Deaths resulting from the primary threat were well-discussed in the observed time period, with consistent discussion taking place from March to May of 2020, followed by a gradual decrease until the winter months. Increase in reports of pandemic-related deaths during the latter part of the observed time period coincide with large increases in weekly death rates following holiday gatherings. 

Messages containing causal narratives of weather events affecting other weather effects often involved larger weather events such as tropical storms and wildfires causing localized phenomena such as flash flooding and poor air quality. These narratives were mostly used during the 2020 summer natural disaster season that brought about large-scale disasters such as Hurricanes Laura and Sally throughout the South and wildfires impacting the Western United States \cite{smith_2020_2021}.

\subsection{Predicting retransmission}
\label{sec:retran}

The results of the negative binomial regression on retransmission are represented in Table ~\ref{rt_reg3}, with results for model controls listed separately in the Appendix (Table~\ref{rt_regs}). As the number of consequences from a causal narrative increases, the likelihood of retransmission slightly decreases (-0.013). However, as an effect is used more throughout the observed time period, retransmission increases (0.139 for every increase in the log cumulative usage of an effect). 

Of the cause themes found to be statistically significant, causes relating to the impacts and spread of COVID, as well as measures taken are more likely to be retransmitted (0.261, 0.596, and 0.329, respectively) compared to causes relating to Secondary Threats. Causes with expressive, emotional appeals are also more likely to be retransmitted (0.365). Causes relating to transitions or shifts from normalcy are less likely to be retransmitted compared to Secondary Threats.

Of the effect themes found to be statistically significant, the following were found to increase retransmission compared to the reference effect theme of Transitions/Shifts: Secondary Threats (0.362), Susceptibility (0.431), Primary Threat Impact (0.307), Primary Threat Measures (0.271), Economic and Financial Support (0.321), and Official Responses (0.501). Emotional Responses and Coping, and Off-Topic messages (mainly pertaining to voting procedures) are found to be strong predictors of retransmission, at 1.116 and 1.145, respectively. Effects related to Data Processing are less likely to be retransmitted compared to effects about Transitions/Shifts, and is the only negative effect that is statistically significant.

\begin{table}[H]
\centering
\begin{tabular}{l|rrr} \hline\hline
&               Estimate &Std. Error &$\Pr(>|z|)$\\
\hline
(Intercept)                              &      -4.656***   &   0.260  &  0.000 \\
Cause In-Degree                          &       0.002\hspace{.55cm}   &   0.007  &  0.741 \\
Effect Out-Degree                        &      -0.013***   &   0.003  &  0.000 \\
Log Follower Count                       &       0.747***   &   0.010  &  0.000 \\
Transitive Closure                       &      -0.113\hspace{.55cm}   &   0.060  &  0.059 \\
Log of Cumulative Cause Usage            &      -0.033\hspace{.55cm}   &   0.025  &  0.180 \\
Log of Cumulative Effect Usage           &       0.139***			   &   0.031  &  0.000 \\
Cause Theme: Primary Threat              &       0.205\hspace{.55cm}   &   0.114  &  0.073 \\
Cause Theme: Susceptibility              &       0.274\hspace{.55cm}   &   0.380  &  0.471 \\
Cause Theme: Primary Threat Impact       &       0.261**\hspace{.15cm}   &   0.094  &  0.005 \\
Cause Theme: Primary Threat Measures     &       0.329*\hspace{.35cm}  &   0.142  &  0.021 \\
Cause Theme: Spread                      &      0.596**\hspace{.15cm}    &  0.187   & 0.001 \\
Cause Theme: Economic and Financial Support &   -0.135\hspace{.55cm}   &   0.126  &  0.285 \\
Cause Theme: Data Processing              &     -0.165\hspace{.55cm}   &   0.194  &  0.394 \\
Cause Theme: Transitions/Shifts           &     -0.319*\hspace{.35cm}   &   0.160  &  0.047 \\
Cause Theme: Official Responses           &     -0.657\hspace{.55cm}   &   0.507  &  0.195 \\
Cause Theme: Emotional Responses and Coping &    0.365**\hspace{.15cm}   &   0.132  &  0.006 \\
Cause Theme: Events/Actors               &       0.064\hspace{.55cm}   &   0.142  &  0.651 \\
Cause Theme: Off-Topic                   &      -0.575\hspace{.55cm}   &   0.629  &  0.360 \\
Effect Theme: Primary Threat             &      -0.063\hspace{.55cm}   &   0.212  &  0.767 \\
Effect Theme: Secondary Threats          &       0.362***   &   0.084  &  0.000 \\
Effect Theme: Susceptibility             &       0.431*\hspace{.35cm}   &   0.192  &  0.025 \\
Effect Theme: Primary Threat Impact      &       0.307***   &   0.072  &  0.000 \\
Effect Theme: Primary Threat Measures    &       0.271**\hspace{.15cm}   &   0.099  &  0.006 \\
Effect Theme: Spread                     &      -0.494\hspace{.55cm}   &   0.370  &  0.182 \\
Effect Theme: Economic and Financial Support &   0.321***   &   0.080  &  0.000 \\
Effect Theme: Data Processing            &      -0.946***   &   1.306  &  0.469 \\
Effect Theme: Official Responses         &       0.501***  &   0.121  &  0.000 \\
Effect Theme: Emotional Responses and Coping &   1.116***   &   0.155  &  0.000 \\
Effect Theme: Off-Topic                  &       1.145***   &   0.343  &  0.001 \\
\hline\hline
\end{tabular}
\caption{\label{rt_reg3} Negative binomial regression results for retweet as outcome variable, as a function of cause in-degree, effect out-degree, transitive closure, and cause and effect themes, while controlling for cumulative cause and effect usage, log follower count, day/hour of posting (shown in the Appendix) and number of months since pandemic started (also shown in the Appendix). Cause Theme coefficients uses the theme of ``Secondary Threats'' as the reference factor, and Effect Theme coefficients use the theme of ``Disruptions'' as the reference factor.  Observations: 6,955; Akaike Information Criterion: 44,600.8; log-likelihhood: -22,239.4; dispersion parameter: 0.573; standard error: 0.01.}
\end{table}

\section{Discussion} \label{sec_discussion}

The semantic causal narrative network constructed from 39 discovered concepts within our corpus reveals that much of the official discourse during the observed time period involved disruptions from normalcy due to either weather events or COVID-19. In evaluating the differences in narrative usage across various organizational roles, we find that a strongly consistent discourse is used during this time period, in which a central focus is on the disruptions to regularly-scheduled events, and on the wide-reaching impacts of COVID-19. Elected officials contrast slightly from responding agencies in their communication patterns, with responding agencies speaking more on the day-to-day impacts of COVID and electeds speaking more on COVID-19 -related assistance. However, we do not find there to be much consistency across time. Causal narrative usage vastly changes across time, with popular narratives changing from a large focus on COVID-19 and its impacts during the first few months of the pandemic in March 2020, to narratives focusing on weather conditions and disasters during the summer and winter disaster seasons.

From our analysis on effectiveness of messages given certain semantic network characteristics, we find that network structures such as causal narrative in-degree and transitive closure are not statistically significant. However, the more consequences a narrative has (effect out-degree), the less likely it is to be retransmitted when accounting for the effect theme. Furthermore, we find that retransmission increases as an effect is attributed to more over time when controlling for time passed from the beginning of 2020 to the time of message dissemination. Fatigue against repeated usage of effects is not found in our analysis. Instead, we suspect that the repeated discussion of a consequence makes it more salient to the public and, thus, more likely to be retransmitted.

In terms of themes, there are certain causes and effects that attract more engagement than others. Many narratives relating directly to COVID-19 such as its impacts and protective measures against it are more likely to garner more engagement than the baseline narrative of weather-related disruptions. One of the strongest predictors of engagement is the theme of emotional appeal, or narratives speaking of resilience, broad challenges, and expressions of gratitude. The results of our analysis suggest that the public is more likely to attend to narratives that provide directed information on the severity of COVID-19 and advise on specific actions to take to safeguard against the threat, and are especially likely to engage with messages in which the actor provides an emotional appeal in response to a situation.

From the results presented here, there is one recurring narrative that was consistently discussed amongst all actors, across time, and often: weather-related disruptions. Although we may expect this to be a narrative that persisted from before the pandemic, there is a marked increase from February of 2020 to February 2021 in discussion of weather-related disruptions. Upon visual inspection, many of these messages refer disruptions to essential services resulting from the pandemic. Many distribution services were moved to outdoor locations during the pandemic to mitigate spread of the virus, including distribution of tests, vaccinations, and blood and food supplies. However, these services then became subject to adverse weather conditions ranging from the everyday rain and storms, to more dire conditions such as hail and freezes.

In addition to the upset of outdoor essential services, disaster response measures were also complicated due to the nature of the pandemic. Among messages from officials were warnings of the threats of disease transmission when evacuating into crowded shelters, such as: ``Your hurricane evacuation plan may need to change due to  \#COVID19.'' Although the included example was in reference to one of the twelve storms to make landfall on the U.S. during the 2020 hurricane season \cite{masters_look_2020}, evacuation procedures were impacted for the numerous disasters impacting the nation including the wildfires of the Western continental U.S. and severe freezes impacting the south. Threatened by these weather events, the disruption of essential services and the complications to evacuation procedures led to even more hazardous conditions for the general public.

\subsection{Future Studies}

Although this work gives an in-depth analysis on causal narratives utilized by organizations during the pandemic, it is limited in scope to causal narratives of three specific conectives (e.g. ``due to,'' ``because,'' and ``caused by'' located between putative causes and effects). Messages containing these causal connectives may only provide a partial view of all the causal narratives used during this time period, and obviously do not speak to messaging that does not employ causal structure. Examples of this may include narratives in which the causal phrases are located at the beginning of the sentence, e.g. ``Due to B, then A,'' which were not included here due to the difficulty of accurate parsing of clauses in these cases. However, the number of messages containing ``Due to..." and "Because of..." in our dataset is quite small, only identified in 2,185 tweets. To understand the extent of this limitation, we conducted a further review of one-thousand randomly sampled messages from the larger corpus, finding 156 instances of causal statements in which 11 of them were captured by our templated method. From this, we estimate that about 154,300 messages contain a causal assertion in our dataset of approximately 989,000 messages and our collection method captured about 7\% of those statements containing a causal assertion. 

Furthermore, although narratives were examined for consistency across actors, further investigation could be made into narrative consistency at the individual account level as well. Although consistency across actors conveys a united front to a public that may not distinguish among different actors, consistencies within a single organization's messages is desirable for public understanding and for building credibility \cite{finset_effective_2020}.  Such investigations are limited by the much smaller number of communications issued by any given account, however.

\subsection{Practical Implications}

As discussed, writing messages in a structured form is beneficial for increased salience and understanding of the message, impacting perception of the risk at hand in crisis settings. Although there was a consistent discourse shared amongst actors within our observed time period, there are obvious advantages to developing coordinated messaging strategies with other organizations. Clear and consistent messaging will lead to optimal conditions for reducing public uncertainty and increasing understanding of actions to take in times of risk. From this study, we find that repeated discussion of a situation's consequences can lead to public amplication of a message by way of retransmission. Messages that contain directed, timely information (for example, the actions one can take to safeguard against the primary threat) will garner increased public engagement, along with messages that contain emotional appeals, such as ones of community resilience and gratitude. 

\section{Conclusion} \label{sec_conclusion}

During the first fifteen months of the pandemic, the COVID-19 pandemic dominated official organizational online discourse, nevertheless sharing that discourse space with regularly occurring issues and crises, such as natural disasters and non-COVID -related illnesses. Much of what was discussed involved disruptions to everyday life due to both the pandemic and weather hazards. Organizations with distinct goals and objectives were nevertheless found to adhere to a common theme, which bodes well for overall effectiveness of messages. There is a lack of consistency in causal narratives across time. However, consistency may have proved difficult to achieve due to the need of organizations to respond to the evolving threat during the  time period. The many non-COVID-19 threat events observed during our study time period are also reminders that in a massive crisis such as a global pandemic, officials and organizations must be prepared to respond to and communicate on matters beyond the primary threat at hand. Despite the devastation of the more salient threat, other crises such as natural disasters will still heavily affect the general public, if not exacerbate and complicate already difficult conditions.

\newpage

\bibliographystyle{apacite}
\bibliography{biblio.bib}

\clearpage
\section*{Appendix}
\begin{table}[]
	\centering
	\begin{tabular}{l|rrr} \hline\hline
		&               Estimate &Std. Error &$\Pr(>|z|)$\\
		\hline
		Period Effects - Day of Week  &&&\\
		\hline
		Monday          &          -0.241**\hspace{.15cm} &   0.08089  & 0.00287 \\
		Tuesday         &          -0.315***  &  0.07940  & 7.4e-05\\
		Wednesday       &            -0.337***  &  0.07928 & 2.1e-05\\
		Thursday        &            -0.329*** &   0.07863 &  2.8e-05\\
		Friday          &       -0.409***   & 0.07935  & 2.5e-07\\
		Saturday        &            -0.323*** &   0.09270 &   0.00048\\
		\hline
		Period Effects - Time of Day  &&&\\
		\hline
		1 AM UTC        &              0.128\hspace{.55cm} &   0.13236 &   0.32999\\
		2 AM UTC        &              0.138\hspace{.55cm} &   0.15653 &   0.37781\\
		3 AM UTC        &              0.362\hspace{.55cm} &   0.20206   & 0.07288\\
		4 AM UTC        &             -0.177\hspace{.55cm} &   0.28208   &  0.53120\\
		5 AM UTC        &             -0.601*\hspace{.35cm} &   0.29486   &  0.04142\\
		6 AM UTC        &             -0.046\hspace{.55cm}  &  0.41611   &  0.91172\\
		7 AM UTC        &             -0.184\hspace{.55cm} &   0.46224   &  0.69120\\
		8 AM UTC        &              0.131\hspace{.55cm} &   0.43755   &  0.76429\\
		9 AM UTC        &            -1.826***  &  0.41643   &  1.2e-05\\
		10 AM UTC       &             -0.797**\hspace{.15cm} &   0.30653   &  0.00929\\
		11 AM UTC       &             -0.044\hspace{.55cm} &   0.19988   &  0.82586\\
		12 PM UTC       &             -0.731*** &   0.13840   &  1.3e-07\\
		1 PM UTC        &            -0.401*** &   0.11623   &  0.00055\\
		2 PM UTC        &            -0.671*** &   0.10634   &  2.8e-10\\
		3 PM UTC        &            -0.326**\hspace{.15cm} &   0.10656   &  0.00224\\
		4 PM UTC        &            -0.360*** &   0.10535   &  0.00062\\
		5 PM UTC        &            -0.266*\hspace{.35cm} &   0.10517   &  0.01145\\
		6 PM UTC        &            -0.427*** &   0.10505   &  4.7e-05\\
		7 PM UTC        &            -0.378*** &  0.10504   &  0.00032\\
		8 PM UTC        &            -0.205\hspace{.55cm} &   0.10609   &  0.05298\\
		9 PM UTC        &            -0.320**\hspace{.15cm} &   0.10700   &  0.00276\\
		10 PM UTC       &             -0.148\hspace{.55cm} &    0.10883  &  0.17430\\
		11 PM UTC       &             -0.059\hspace{.55cm} &   0.11539  &  0.61123\\
		\hline
		Num. of Months   &            -0.082*** &   0.00738  &  $<$ 2e-16\\
		\hline\hline
	\end{tabular}
	\caption{\label{rt_regs} Negative binomial regression results for controls. ``Num. of Months" is the number of months into the observation period (January 2020 to March 2021, with January 2020 being Month 1.) Observations: 6,955; Akaike Information Criterion: 44,600.8; log-likelihhood: -22,239.4; dispersion parameter: 0.573; standard error: 0.01.}
\end{table}

\end{document}